\documentclass[journal,twoside,web]{IEEEtran}
\usepackage{cite}
\usepackage{amsmath,amssymb,amsfonts}
\usepackage{algorithmic}
\usepackage{graphicx}
\usepackage{textcomp}
\usepackage{bm}
\usepackage{makecell}
\usepackage{multirow}
\usepackage{color}
\usepackage[caption=false]{subfig}
\usepackage{bbding}
\usepackage{amssymb}
\usepackage{pdfrender}
\newcommand*{\boldtimes}{%
  \textpdfrender{
    TextRenderingMode=FillStroke,
    LineWidth=.5pt, 
  }{\times}%
}
\def\BibTeX{{\rm B\kern-.05em{\sc i\kern-.025em b}\kern-.08em
    T\kern-.1667em\lower.7ex\hbox{E}\kern-.125emX}}
\begin{document}
\title{Sequence-to-Sequence Voice Reconstruction for Silent Speech in a Tonal Language}
\author{Huiyan~Li,
        Haohong~Lin,
        You~Wang,
        Hengyang~Wang,
        Ming~Zhang,\\
        Han~Gao,
        Qing~Ai,
        Zhiyuan Luo,
        and Guang~Li
\thanks{This work is supported by the Science Foundation of Chinese Aerospace Industry under Grant JCKY2018204B053 and the Autonomous Research Project of the State Key Laboratory of Industrial Control Technology, China(Grant No. ICT2021A13).}
\thanks{H.~Li,~
        H.~Lin,~
        Y.~Wang,~
        H.~Wang,~
        M.~Zhang,~
        H.~Gao,~
        Q.~Ai 
        and G.~Li are with the State Key Laboratory of Industrial Control Technology, Institute of Cyber Systems and Control, Zhejiang University, Hangzhou 310027, China (e-mail: \{huiyanli, lhh2017, king\_wy, 11432014, drystan, gao\_han, aiqing, guangli\}@zju.edu.cn).}
\thanks{Z.~Luo is with Department of Computer Science, Royal Holloway, University of London,  Egham Hill, Egham, Surrey TW20 0EX, UK (e-mail: Zhiyuan.Luo@cs.rhul).}
}

\maketitle

\begin{abstract}


Silent Speech Decoding (SSD), based on articulatory neuromuscular activities, has become a prevalent task of Brain-Computer Interface (BCI) in recent years. Many works have been devoted to decoding surface electromyography (sEMG) from articulatory neuromuscular activities. However, restoring silent speech in tonal languages such as Mandarin Chinese is still difficult. This paper proposes an optimized Sequence-to-Sequence (Seq2Seq) approach to synthesize voice from the sEMG-based silent speech. We extract duration information to regulate the sEMG-based silent speech using the audio length. Then, we provide a deep-learning model with an encoder-decoder structure and a state-of-art vocoder to generate the audio waveform. Experiments based on six Mandarin Chinese speakers demonstrate that the proposed model can successfully decode silent speech in Mandarin Chinese and achieve a character error rate (CER) of 6.41\% on average with human evaluation.

\end{abstract}


\section{Introduction}
%
%
%
%


Silent Speech Decoding (SSD) is one of the most popular areas of Brain-Computer Interface (BCI) research, which makes it possible for humans to interact with their surroundings and express their inner minds without speaking words~\cite{denby2010silent, wang2020silent}. SSD aims at detecting biological speech-related activities (instead of acoustic data) and decoding the thought of humans using physiological measurements. 

Speech-related signals detected by physiological measurements are defined as \textit{biosignals}~\cite{DBLP:journals/taslp/SchultzWHKHB17}. The typical physiological measurements are obtained by using sensors to capture biosignals from brain~\cite{DBLP:journals/access/LopezAMPG20}, e.g., electrocorticography (ECoG)~\cite{10.3389/fnins.2015.00217, DBLP:journals/ijon/AngrickHJSKS19, angrick2019speech}, and electroencephalography (EEG)~\cite{DBLP:journals/ijon/RamadanV17, DBLP:conf/biostec/PorbadnigkWCS09}.  However, these devices for biosignals acquisition have several disadvantages. ECoG is invasive and probably has surgical complications~\cite{rolston2016major}; EEG has no harmful side effects, but the signal processing of EEG is difficult for practical use~\cite{wang2020silent}. 
Acquisition of neuromuscular signals is a promising way to decode speech-related activity~\cite{DBLP:journals/taslp/SchultzWHKHB17}. 


Surface electromyography (sEMG), which is non-invasive and convenient to apply in practical applications, can be used to acquire the control signals that are transferred from the cortex to the facial muscles and then decode the silent speech~\cite{DBLP:conf/ijcnn/DienerJS15}. In addition, the neural pathways from the brain to muscle can act as primary filters and encoders~\cite{DBLP:journals/tbe/WandJS14}, EMG has lower channel requirements~\cite{wang2020silent}. Electromagnetic Articulography (EMA) sensors~\cite{fagan2008development}, and optical imaging of the tongue and lips~\cite{DBLP:conf/icassp/DenbyS04} are also often used in SSD to record invisible speech articulators. However, they could not work in the absence of articulator movement.

Existing studies on SSD can be divided into two categories: \textit{biosignal-to-text} and \textit{biosignal-to-voice}~\cite{DBLP:journals/taslp/SchultzWHKHB17}. 
The former one can be regarded as a kind of classification task, while the latter one is a kind of regression task~\cite{DBLP:journals/ijon/AngrickHJSKS19}. 
Considering that the biosignal-to-text approaches may lose some information about the speaker's personality and emotion during processing, the two-step approach ``biosignal-to-text-to-voice" is too time-consuming for real-time scenarios~\cite{angrick2019speech}. And many works have tried to decode silent speech by reconstructing voices~\cite{angrick2019speech, DBLP:journals/taslp/SchultzWHKHB17}. In this paper, we denote this task as sEMG-to-voice ($sEMG2V$). This technology has many applications. It will not be interfered by external noise, which makes it remain effective in noisy environments such as factories. This technology can also help patients who are no longer able to speak due to surgical removal of their larynx caused by trauma or diseases~\cite{DBLP:journals/speech/DenbySHHGB10, DBLP:journals/taslp/JankeD17, DBLP:conf/emnlp/GaddyK20, DBLP:journals/taslp/MeltznerHDLRK17}. Besides, this mode is more concealed and can not be observed through lip language analysis and other means, which offer more privacy protection.

The existing methods of $sEMG2V$ in a tonal language have the following problems:

Research on SSD is mainly concentrated in non-tonal languages such as English~\cite{DBLP:journals/taslp/JankeD17, DBLP:conf/emnlp/GaddyK20} while SSD approaches for tonal languages are limited to solving the tasks of classification~\cite{zhang2020inductive, wang2020silent, DBLP:journals/ijon/WangZWWLL21}. Different from non-tonal languages, the pitches, called tones in tonal languages, carry more lexical or grammatical information to distinguish one word from another~\cite{li2021human, kaan2013changes, huang2022just, chen2022computational}. It has been shown that tones carry no less information than vowels in Mandarin Chinese~\cite{DBLP:conf/interspeech/SurendranLX05}. The distinctive tonal patterns of language are called tonemes~\cite{yip2002tone, DBLP:conf/icassp/LeiJNBO05} to distinguish from phonemes. There are five tones in Mandarin Chinese, which are transcribed by letters with diacritics over vowels~\cite{trask2004dictionary}: high level tone (first tone) like /b\={a}/ (eight), rising tone (second tone) like /b\'{a}/ (to pull), dipping tone (third tone) like /b\v{a}/ (to hold), high-falling tone (fourth tone) like /b\`{a}/ (father), and neutral tone (fifth tone) like /ba/ (an interrogative particle). The total number of tonemes, including toned vowels and consonants, is 139 in Mandarin Chinese task~\cite{{DBLP:conf/lrec/SchultzS14}}, while the total number of phonemes in non-tonal languages such as English is 47~\cite{DBLP:journals/corr/abs-2106-01933}. With the same number of datasets, Mandarin contains a larger dimension of information than English and is more difficult to decode. As a result, SSD approaches for tonal languages are limited to solving the classification task~\cite{zhang2020inductive, wang2020silent, DBLP:journals/ijon/WangZWWLL21}. One research recognizes 10 Chinese words in silent speech with an accuracy of 90\%~\cite{wang2020silent}. To our knowledge, there is few work studying on sEMG2V in tonal languages.

In addition, the sEMG-based silent speech has no time-aligned parallel audio. To provide time-aligned parallel information, dynamic time warping (DTW)~\cite{DBLP:conf/kdd/BerndtC94} can be applied to get alignments between silent and vocal speech~\cite{DBLP:journals/taslp/JankeD17}. Recently, Gaddy and Klein~\cite{DBLP:conf/emnlp/GaddyK20} utilize predicted audio for DTW to achieve alignments, extract audio features in parallel, and obtain a word error rate (WER) of 68.0\% in the sEMG-based silent speech. The accuracy can still be improved by finding a better approach for providing corresponding audios.

In order to address these limitations of existing $sEMG2V$ methods in the tonal language, this paper proposes a novel approach based on a Sequence-to-Sequence (Seq2Seq) model, inspired by the tremendous success of the Seq2Seq model in text-to-speech (TTS) and voice conversion (VC)~\cite{DBLP:conf/icassp/HayashiHKT21,DBLP:conf/nips/RenRTQZZL19, DBLP:conf/iclr/0006H0QZZL21, DBLP:conf/nips/KimKKY20}. This technology can help solve the unparallel information between silent speech and audio. We utilize length regulator~\cite{DBLP:conf/nips/RenRTQZZL19} for the $sEMG2V$ to obtain audio signals. The key contributions of this paper are summarized as follows:
\begin{enumerate}
\item The paper proposes a Seq2Seq model, the first attempt to introduce a Seq2Seq model into the $sEMG2V$ task. The model extracts duration information from the alignment between sEMG-based silent speech and vocal speech. The lengths of input sequences are adjusted to match the size of output sequences. Thus, our model can generate audios from neuromuscular activities using the Seq2Seq model.
\item The model in the paper generates audios from sEMG-based silent speech by considering both vocal sEMG reconstruction loss and toneme classification loss, and uses a state-of-art vocoder to achieve better quality and higher accuracy of the reconstructed audios.
\item We collect an sEMG-based silent speech dataset with Mandarin Chinese and conduct extensive experiments to demonstrate that the proposed model can decode neuromuscular signals in silent speech successfully in the tonal language. 
\end{enumerate}

\begin{table}[!t]
\caption{Electrode Locations Details}
\centerline{
\begin{tabular}{ccc}
\hline
\multicolumn{1}{c}{\makecell[c]{Electrode\\Index}} & \makecell[c]{Position} \\
\hline
\multicolumn{1}{c}{1} 

&\multicolumn{1}{c}{\makecell{1 cm right\\from the nose}} \\
\hline

\multicolumn{1}{c}{2}&{\makecell{1 cm right from\\corners of the mouth}}\\
\hline

\multicolumn{1}{c}{3}&{\makecell{1 cm left from the nose}}\\
\hline
\multicolumn{1}{c}{4} &\multicolumn{1}{c}{\makecell{left corner of chin}}\\
\hline
\multicolumn{1}{c}{5} &\multicolumn{1}{c}{\makecell{4 cm behind the chin}}\\
\hline
\end{tabular}}
\label{tab:electrodes}
\end{table}

\begin{figure}[!t]
\centerline{
\subfloat[Main view]{\includegraphics[width=1.1in]{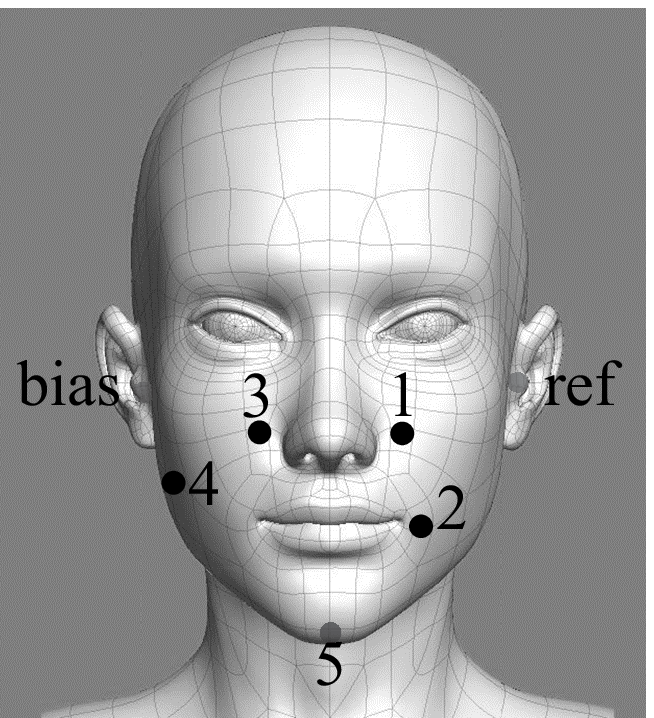}}
\subfloat[Right view]{\includegraphics[width=1.1in]{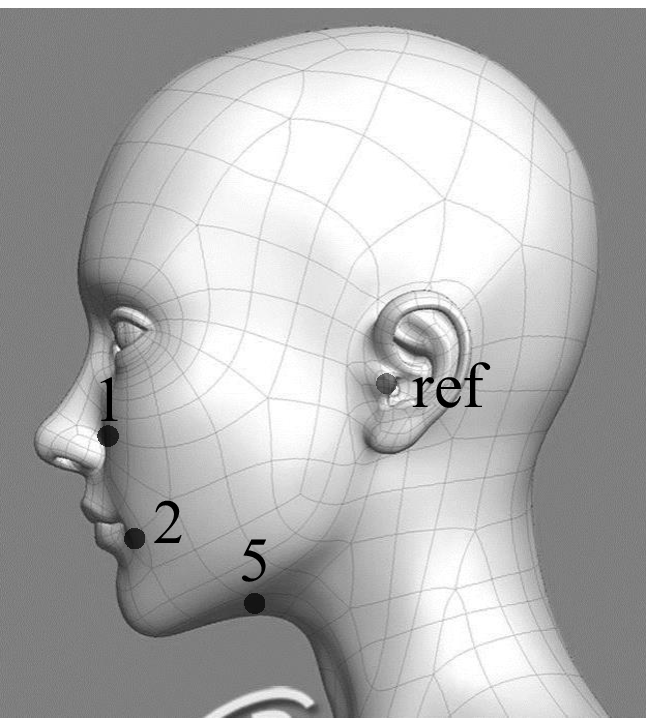}}
\subfloat[Left view]{\includegraphics[width=1.1in]{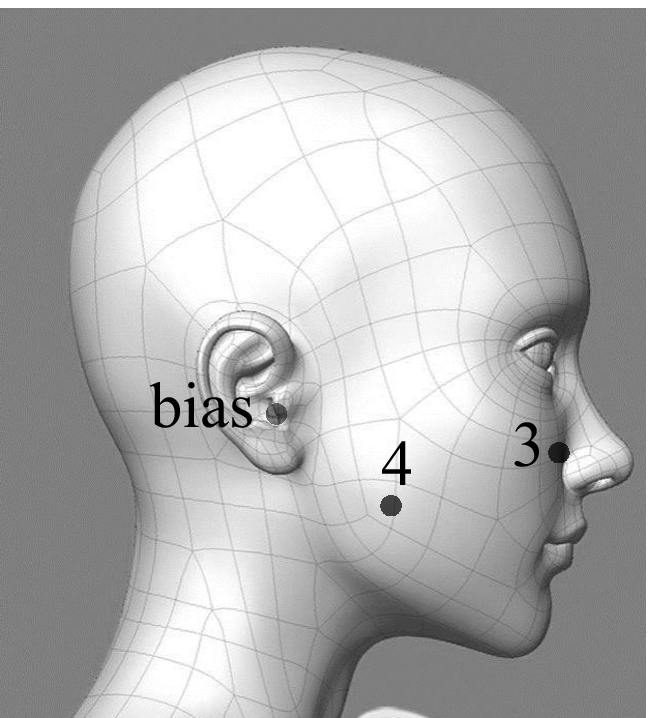}}}
\caption{Three views of electrodes distribution around the face and neck. }
\label{fig:electrodes}
\end{figure}

\begin{table}[t]
\caption{Statistics of the $sEMG\_Mandarin$ Dataset}
\label{experiment_details}
\centerline{
\begin{tabular}{cccccccc}
\hline
\multirow{2}{*}{\makecell[c]{Speaker\\id}} & \multirow{2}{*}{Sex} & \multicolumn{3}{l}{Silent Speech Time (Minutes)} & \multicolumn{3}{l}{Number of utterances} \\ \cline{3-8} 
 &  & Train & Val & Test & Train & Val & Test \\\hline
1 & f & 52.59 & 6.68 & 6.49 & 680 & 85 & 85 \\
2 & m & 52.65 & 6.70 & 6.26 & 516 & 64 & 64 \\
3 & f & 56.60 & 7.11 & 6.99 & 716 & 89 & 89 \\
4 & m & 40.77 & 5.11 & 5.01 & 800 & 100 & 100 \\
5 & m & 40.49 & 5.06 & 5.09 & 600 & 75 & 75 \\
6 & m & 34.85 & 4.29 & 4.40 & 600 & 75 & 75 \\\hline
\multicolumn{2}{c}{Total} & 277.95 & 34.95 & 34.25 & 3912 & 488 & 488\\\hline

\end{tabular}}
\end{table}

\section{Data Acquisition}\label{sec:data}
The acquisition of experimental data is discussed in this section. We first detail the recording information and then describe the dataset information. Finally, we present signal conditioning and feature extraction.

\subsection{Recording Information}

The signal from facial skin is collected by a multi-channel sEMG data recording system using standard wet surface Ag/AgCl electrodes, as described in~\cite{wang2020silent}. Meanwhile, we use a headset microphone to record audio. The views of the electrodes around the face are shown in Fig.~\ref{fig:electrodes}, and the electrode positions are shown in Table~\ref{tab:electrodes}. The electrode positions are highly correlated with vocalizing muscles and have different meanings in speech production~\cite{DBLP:journals/ijon/WangZWWLL21}. In our case, channel 1 is differential electrodes, and the others are single electrodes. The differential electrodes can improve the common-mode rejection ratio and improve the quality of signal~\cite{zhang2020inductive}.

\subsection{Dataset Information}

We collect the data from six native Mandarin-speaking healthy young Chinese adults with normal vision and oral expression skills. The average age of the six participants is 25. The participants are asked to clean their face before the experiment and sit still wearing electrodes and a microphone. They are trained to press the start button, read the sentences shown on the computer screen in vocal and silent mode and press the end button. In silent mode, the participants are trained to imagine speaking sentences displayed on the computer screen as~\cite{wang2020silent} shows. The dataset includes the pair of simultaneously recorded vocal sEMG ($sEMG_{v}$) and audio signal ($Audio_{v}$), and silent sEMG data (namely, $sEMG_{s}$). 
The vocal mode is recorded once, while the silent mode is repeated five times. 
Each recording uses phonetically balanced utterances from a Chinese Corpus called AISHELL3~\cite{DBLP:journals/corr/abs-2010-11567}. There are a total of 2260 words and 1373 characters in this dataset. The dataset includes six speakers, and each of them has at least 0.73 hours of silent speech data, leading to 5.79 hours in total. The dataset of each speaker is split into a training, validation, and testing set, with a ratio of 8: 1: 1 according to the number of silent utterances from each speaker, ensuring that they are phonetically balanced. Table~\ref{experiment_details} gives some statistics of each speakers. In the following, the collected dataset is denoted as $sEMG\_Mandarin$.

\begin{figure}[!t]
\centerline{\includegraphics[width=3.3in]{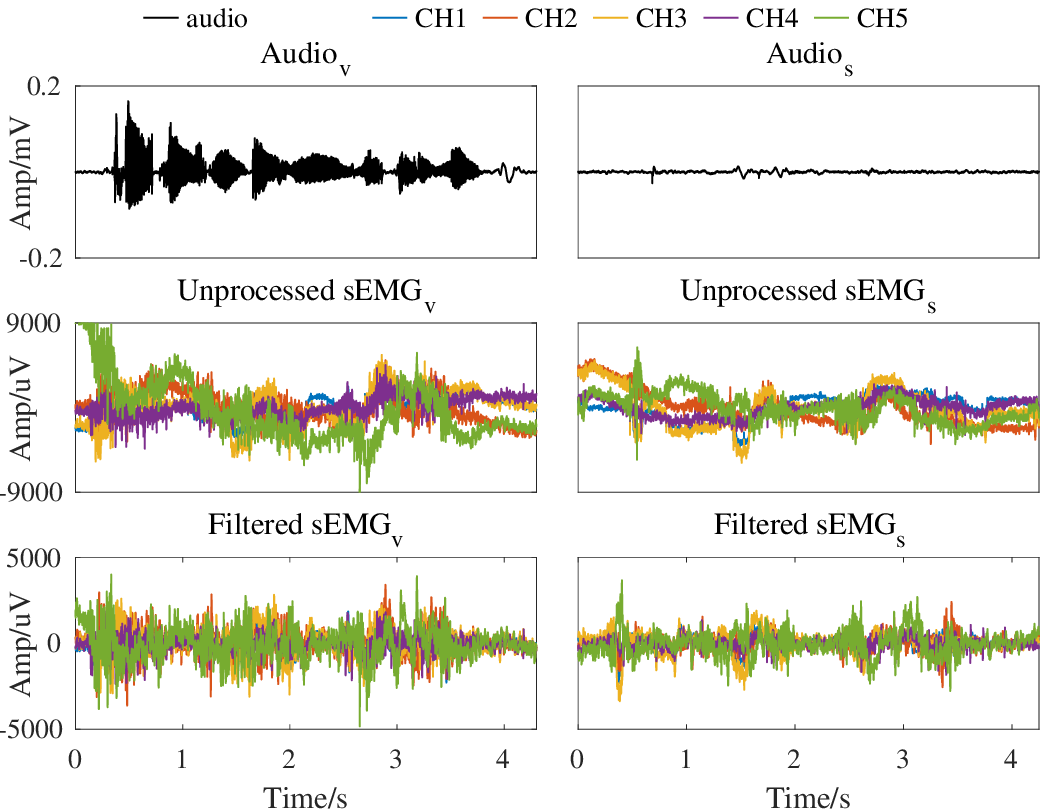}}
\caption{Time-series plots of the audio waveform and neuromuscular signal from Spk-3 with vocal mode and silent mode for the Chinese sentence ``tui4 yi4 yun4 dong4 yuan2 you3 shen2 me5". The audio and neuromuscular signal of the same mode are collected time-synchronized.}
\label{fig:signal}
\end{figure}
\subsection{Signal Conditioning}

The experimental system captures five channels of the sEMG with a sampling frequency of 2000 Hz. 
Butterworth bandpass filter (4 $\sim$ 400 Hz) is applied to remove the offset and high frequency of the signal. A self-tuning notch filter is used to remove the power frequency of 50 Hz and its harmonics~\cite{benesty2006speech}. Audio is recorded with a sampling frequency of 16 kHz. One example of the collected signals with the audio and their five-channel sEMG signals are presented in Fig.~\ref{fig:signal}.

\subsection{Feature Extraction}
To extract the feature of sEMG, we use time-domain (TD) features and time-frequency domain features from the amplitude of short-time Fourier transform (STFT), with 64 ms Hanning window and 16 ms hop length~\cite{DBLP:conf/emnlp/GaddyK20, wang2020silent}. Six TD features are calculated from one frame followed by \cite{jou2006towards}. Finally, 5*6-dimensional TD features and 5*65-dimensional STFT features are extracted and concentrated, i.e., 355-dimensional features are used as input to our model.

To maintain the alignment with sEMG, we extract an 80-dimensional mel-spectrogram with the band-limited frequency range (80 $\sim$ 7600 Hz) from $Audio_{v}$, in which the window length is 1024 points and the hop length is 256 points~\cite{DBLP:conf/icassp/YamamotoSK20}.








\section{The Proposed Methods}\label{sec:model}
We first introduce the target task and overview the proposed model, and then we detail each module of the proposed model.

\subsection{Overview}

\begin{figure}[!t]
\centerline{\includegraphics[width=3.3in]{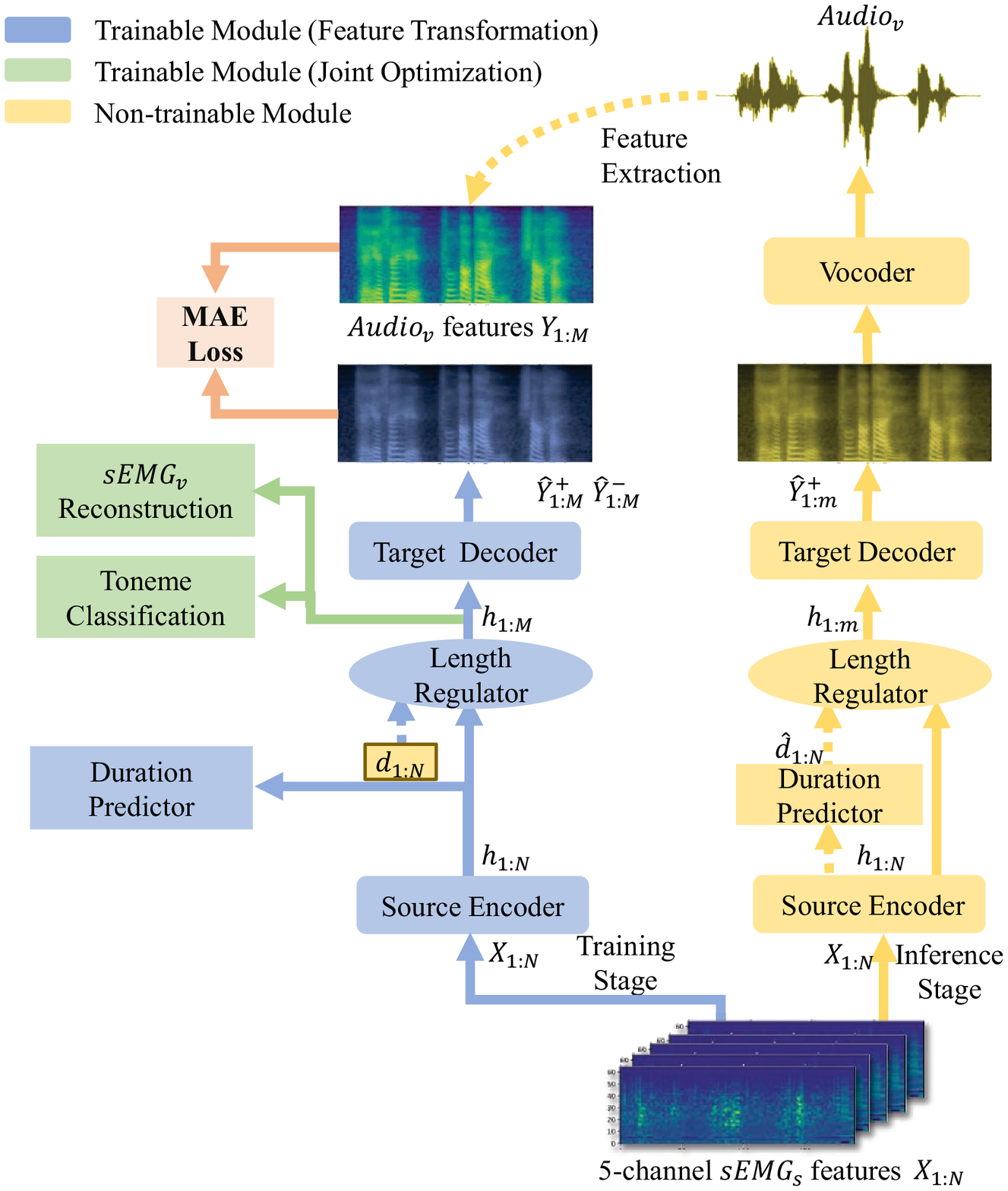}}
\caption{
The overview of the training and inference stages in the SSRNet model. Blue and green blocks represent the feature transformation and joint optimization of the training module, respectively. Yellow blocks represent the no-trainable module, using a pre-trained model to predict the mel-spectrograms without the joint optimization part. We will detail the duration predictor in Section~\ref{sec:c}, and then detail $sEMG_{v}$ Reconstruction Module and Toneme Classification Module in Section~\ref{sec:d}.}
\label{fig:model}
\end{figure}

\begin{figure*}[ht]
\subfloat[Source encoder]{
     \label{fig:encoder}
    \includegraphics[width=6.9in]{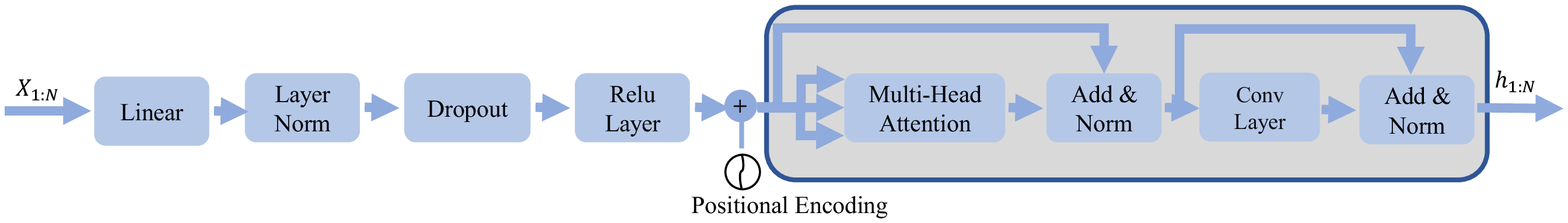}
    }\\
\subfloat[Length regulator]{
     \label{fig:lr}
    \includegraphics[width=1.33in]{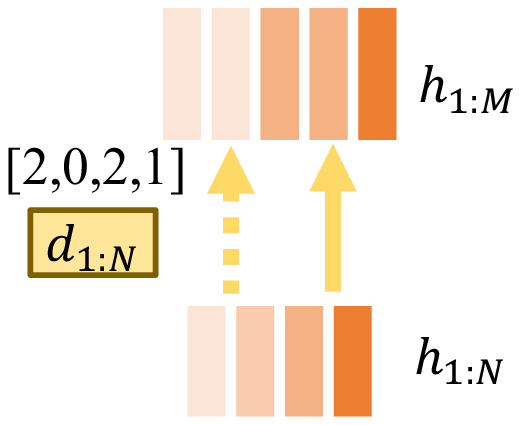}
}
\subfloat[Target decoder]{
    \label{fig:decoder}
    \includegraphics[width=5.65in]{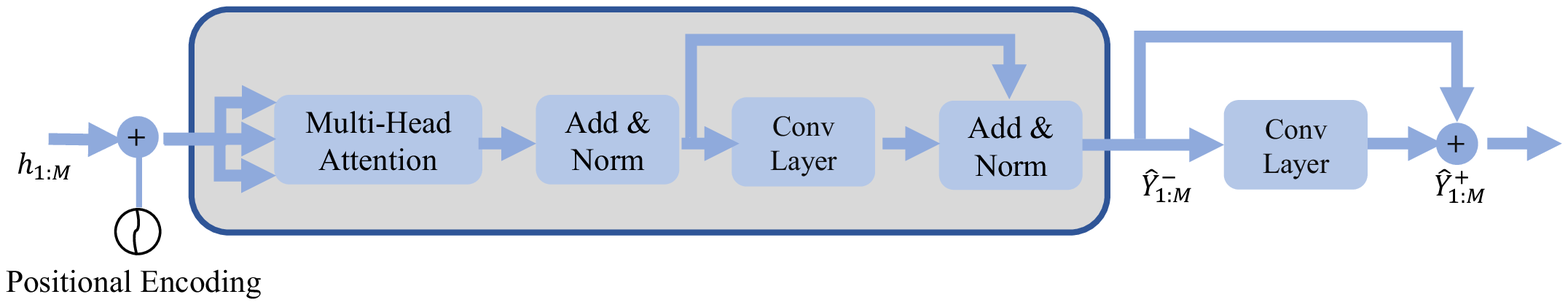}%
    }
\caption{Illustrations of model details about the source encoder, length regulator and the target decoder. ${{\hat Y}_{1: M}^{+}}$ and ${{\hat Y}_{1: M}}^{-}$ is the mel-spectrograms predicted before and after the postnet. The grey blocks represent FFT module.}
\label{fig:model_detail}
\end{figure*}

In order to distinguish between two kinds of sEMG modes, ${X}_{1: N}$ represents $sEMG_{s}$ features while ${x}_{1: M}$ represents $sEMG_{v}$ features. Besides, ${Y}_{1: M}$ represents the mel-spectrograms from $Audio_{v}$.
The target task, i.e., the goal of the $sEMG2V$, is essentially to transform an $N$-length time-series sequence ${X}_{1: N}$ into an $M$-length sequence ${Y}_{1: M}$. Note that the length $M$ of the target sequence ${Y}_{1: M}$ is unknown and depends on the source sequence itself. 

To fulfil this task, we design a novel $sEMG2V$ model, called \textbf{S}ilent \textbf{S}peech \textbf{R}econstruction \textbf{N}etwork (SSRNet in short), see Fig.~\ref{fig:model}. SSRNet generates the mel-spectrograms ${Y}_{1: M}$ directly from the features of $sEMG_{s}$ ${X}_{1: N}$. Moreover, SSRNet resamples the input sequence according to the duration sequence ${d}_{1: N}$ (i.e., ${X}_{1:N}[i]$, the index $i$ of ${X}_{N}$ corresponds to $\left[Y_{1: M}[j], \cdots, Y_{1: M}[j+p-1]\right]$ of ${Y}_{M}$, where $p$ is the duration of ${X}_{1:N}[i]$, called $d_{1:N}[i]$) which is calculated from the alignment between $sEMG_{s}$ and $Audio_{v}$. Finally, SSRNet transfers the predicted mel-spectrograms ${\hat Y}^{+}_{1: M}$ to the audio waveform by a pre-trained vocoder. 

The procedure mentioned above can be formally described as follows:

\begin{equation}
h_{1: N}=\text{Encoder}\left(X_{1: N}\right)\label{con:1}
\end{equation}
where ${h}_{1: N}$ is the hidden representations produced by the source encoder.
\begin{equation}
h_{1: M}=\text { LengthRegulator }(h_{1: N},d_{1:N})\label{con:2}
\end{equation}
where ${h}_{1: M}$ is generated from ${h}_{1: N}$ by the length regulator, note that $M=\sum_{i=1}^{N} d_{1:N}[i]$ and ${d}_{1: N}$ is the Ground-Truth duration (GT duration) after the alignment.
\begin{equation}
{\hat Y}^{+}_{1: M}=\text {Decoder}\left(h_{1: M}\right)\label{con:3}
\end{equation}
where ${\hat Y}^{+}_{1: M}$ is the mel-spectrograms predicted by the decoder.
\begin{equation}
Audio =\text {Vocoder}\left({\hat Y}^{+}_{1: M}\right)\label{con:4}
\end{equation}


In the inference stage, we use modules of the feature transformation and the duration predicted by the duration predictor instead of the GT duration. The inference stage is also illustrated in Fig.~\ref{fig:model}, ${h}_{1: m}$ is the same as ${h}_{1: M}$; 
${\hat d}_{1: N}$ is the predicted duration by the duration predictor; and ${\hat Y}^{+}_{1: m}$ represent the mel-spectrograms predicted in the inference module, where $m=\sum_{i=1}^{N}\hat d_{1:N}[i]$.


\begin{figure*}[!t]
\centerline{
\includegraphics[width=6.9in]{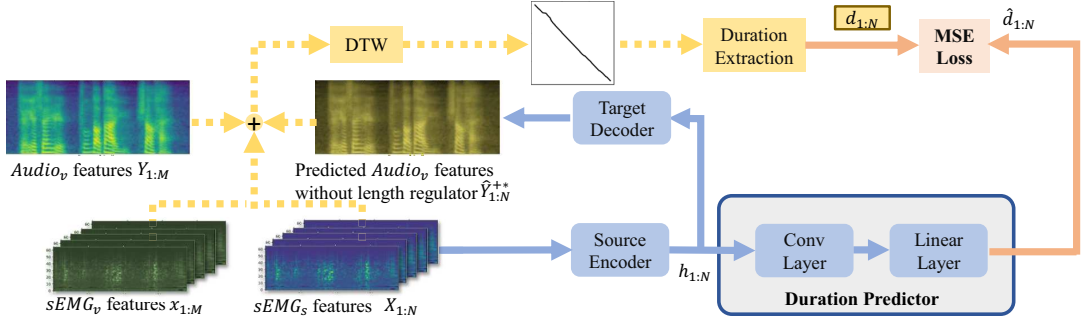}}
\caption{Illustration of duration extraction and predictor. The grey block represents the duration predictor.}
\label{model_lr_new}
\end{figure*}

\subsection{Feature Transformation\label{sec:b}}
The feature transformation module aims to transform the sEMG features to audio features by the length regulator using GT duration. The architecture for the feature transformation in SSRNet, includes an encoder, a length regulator, and a decoder. The main structure of SSRNet is called Feed-Foward Transformer (FFT)~\cite{DBLP:conf/nips/VaswaniSPUJGKP17}, which consists of self-attention in Transformer and 1D convolution layers. FFT aims at exploring the relationship between ${X}_{1: N}$ and ${Y}_{1: M}$ at different positions. This module follows the setting in~\cite{DBLP:conf/nips/RenRTQZZL19}.

The source encoder block, illustrated in Fig.~\ref{fig:model_detail}(a), uses a fully-connected layer with Rectified Linear Units (Relu) activation to convert multi-dimension features of the sEMG to match the FFT hidden size~\cite{DBLP:conf/interspeech/HuangHWKT20}. The positional encoding is introduced to concatenate with the output of the linear layer in order to highlight the position of the frame in ${X}_{1: N}$. After that, SSRNet uses a multiple FFT structure (shown as grey blocks in Fig.~\ref{fig:model_detail}(a) and~\ref{fig:model_detail}(c)) with multi-head attention and a two-layer 1D convolutional network. 

SSRNet applies a length regulator to adjust the length of output hidden representations of the source encoder block to match the output features. Fig.~\ref{fig:model_detail}(b) depicts the length regulator where the length of the input is four, while the length of the output is five. The length of the regulated sequence is adjusted to five by the GT duration ${d}_{1: N}$. The duration from the alignment between ${X}_{1: N}$ and ${x}_{1: M}$ is denoted as the GT duration which will be detailed in Section~\ref{sec:c}. Note that ${d}_{1: N}$ is only used in the training procedure. In the inference procedure, we use the output ${\hat d}_{1: N}$ from the duration predictor as duration to regulate.

The FFT layer used by the target decoder block is the same as the source encoder. As illustrated in Fig.~\ref{fig:model_detail}(c), the output hidden representations after FFT blocks are passed through the linear layer. Mel-spectrograms predicted after the linear layer is ${{\hat Y}_{1: M}^{-}}$. SSRNet further uses convolutional layers called postnet to calculate the residual of the predicted mel-spectrograms, which is used to improve the reconstruction ability of the model~\cite{DBLP:conf/icassp/ShenPWSJYCZWRSA18}. ${{\hat Y}_{1: M}^{+}}$ is the sum of ${{\hat Y}_{1: M}^{-}}$ and the residual mel-spectrograms.

In the feature transformation, SSRNet uses the Mean Absolute Deviations Error (MAE) as the loss function. To be more specific, we minimize the summed MAE of between ${{\hat Y}_{1: M}^{+}}$ and ${Y}_{1: M}$, and between ${{\hat Y}_{1: M}^{-}}$ and ${Y}_{1: M}$.

\subsection{Duration Extractor\label{sec:c}}
Given the synchronization between ${X}_{1: N}$ and ${x}_{1: M}$, the duration extractor uses dynamic programming to achieve pairs position between N-length ${X}_{1: N}$ and M-length ${x}_{1: M}$~\cite{DBLP:conf/icassp/DesaiRYBP09, DBLP:conf/emnlp/GaddyK20}. The cost function is defined as follows: 

\begin{equation}
\left\|X_{1: N}[i]-x_{1: M}[j]\right\|, \ 1 \leq i \leq N, 1 \leq j \leq M\label{eq:align_old}
\end{equation}

Besides, similar as the predicted audio refinement~\cite{DBLP:conf/emnlp/GaddyK20}, the model without the length regulator gets $N$-length predicted audio features ${\hat Y}^{+*}_{1: N}$ during the training procedure. As illustrated in Fig.~\ref{model_lr_new}, the new cost function for DTW in this method is shown as follows:

\begin{equation}
\left\|X_{1: N}[i]-x_{1: M}[j]\right\|+\lambda_{{align }}\left\|\hat{Y}_{1: N}^{+*}[i]-Y_{1: M}[j]\right\|\label{eq:align}
\end{equation}
where $\lambda_{{align}}$ is the weight of audio alignments. 

Instead of achieving a warped audio sequence from pairs, the proposed SSRNet model calculates the duration sequence from the pairs as follows:

\begin{equation}
\ d_{1:N}[i]=\sum_{j=1}^{M}(A_{1:M}[j]==i) \label{eq:dura}
\end{equation}
where $A_{1:M}$ is a length of M sequence which represents whether the index $i$ of the input features is corresponding to the index $j$ of the output features.




Duration predictor aims at predicting the length of audio features corresponding to each frame of sEMG features.
The duration extractor is based on the DTW algorithms~\cite{DBLP:conf/kdd/BerndtC94}. SSRNet trains a duration predictor (i.e.,  convolutional layers and a linear layer) and uses Mean Square Error (MSE) to calculate the loss between GT duration ${d}_{1: N}$ and the predicted duration ${\hat d}_{1: N}$.

\begin{figure}[!t]
\centerline{
\includegraphics[width=3.3in]{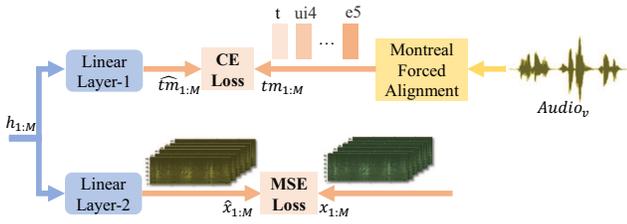}}
\caption{Detail of joint optimization. ${tm}_{1: M}$ is the tonemes with tones from MFA while ${\hat tm}_{1: M}$ is the tonemes predicted after the linear layer. ${x}_{1: M}$ is the $sEMG_{v}$ while ${\hat x}_{1: M}$ is the $sEMG_{v}$ reconstructed from the linear layer.}
\label{model_side}
\end{figure}

\subsection{Joint Optimization with Toneme Prediction and Vocal sEMG Reconstuction\label{sec:d}}
The module of joint optimization with toneme prediction and vocal sEMG reconstuction aims at improving the model performance.
SSRNet employs the pre-trained Mandarin model from Montreal Forced Alignment (MFA) to get the tonemes alignment $tm_{1: M}$ of the audio~\cite{DBLP:conf/interspeech/McAuliffeSM0S17, DBLP:journals/corr/WangZ15e}. The set of toneme for Mandarin is created by GlobalPhone~\cite{DBLP:conf/lrec/SchultzS14} by splitting into onset, nucleus (any vowel sequence), and codas, and then associating the tone of the syllable onto the nucleus (i.e., /teng2/  is split as /t e2 ng/. In Fig.~\ref{model_side}, the hidden representations pass to a linear layer to predict a sequence (including silent frames) ${\hat tm}_{1: M}$, and SSRNet uses the Cross-Entropy (CE) to measure the loss between the target and the output. The purpose of module is to conserve information of the target context. 

Besides, another linear layer at the same position in Fig.~\ref{model_side} is used to restore the hidden representation to the $sEMG_{v}$ for the stable training procedure.

During inference stage, the joint optimization module is discarded.
The joint loss function of the proposed SSRNet model is formulated as follows:

\begin{equation}
\begin{aligned}
\mathcal{L}_{{SSRNet}} &=MAE({\hat Y}^{+}_{1: M},\ Y_{1: M}) \\
&+MAE({\hat Y}^{-}_{1: M},\ Y_{1: M}) \\
&+MSE(\hat{d}_{1: N},\ d_{1: N}) \\
&+\lambda_{{tm}}CE(\hat{tm}_{1: M},\ tm_{1: M}) \\
&+\lambda_{{recons}}MSE(\hat{x}_{1: M},\ x_{1: M})
\end{aligned}
\label{eq:loss}
\end{equation}
where $\lambda_{{tm}}$ controls the toneme classification loss and $\lambda_{{recons}}$ controls the $sEMG_{v}$ reconstruction loss.

\subsection{Vocoder\label{sec:e}}

This paper utilizes Parallel WaveGAN (PWG) as the final synthesizer of desired audible speech~\cite{DBLP:conf/icassp/YamamotoSK20}. This vocoder is an upgraded non-autoregressive version of WaveNet model~\cite{DBLP:conf/ssw/OordDZSVGKSK16}. Unlike some previous non-autoregressive methods such as~\cite{prenger2019waveglow, peng2020non, song2019excitnet}, PWG gets rid of the teacher-student framework, which significantly facilitates our training process and speeds up in the inference stage. 

To synthesize natural $Audio_{v}$, PWG requires an input of auxiliary features, which is ${Y}_{1:M}$ for training and $\hat{Y}_{1:M}$ for inference. The model consists of a non-autoregressive WaveNet generator and a discriminator with non-causal dilated convolution. Instead of the traditional sequential teacher-student framework, PWG has a structure of generative adversarial network (GAN) and jointly optimizes adversarial function loss $L_{adv}$ and the auxiliary loss $L_{aux}$ of multi-resolution STFT loss~\cite{DBLP:conf/ssw/OordDZSVGKSK16}. The loss function of the multi-tasking generator is defined as:
\begin{figure}[!t]
\centering
\includegraphics[width=3.3in]{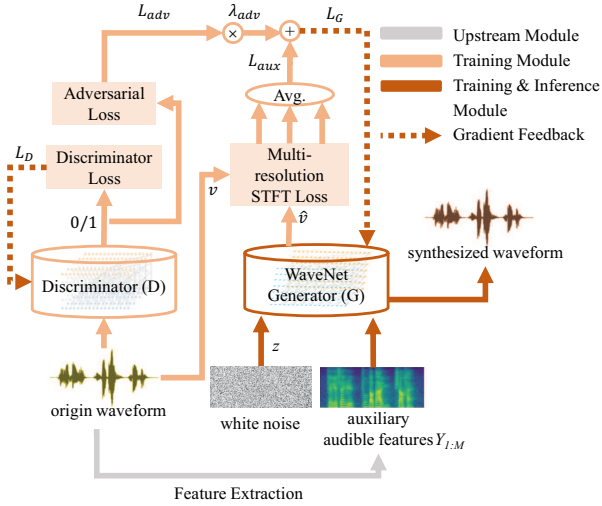}
\caption{The framework of PWG. Orange blocks represent the training module, while red blocks represent both the training and inference module. Dotted lines mean gradient feedback.}
\label{fig:PWG}
\end{figure}

\begin{equation}
\begin{aligned}
&L_{G}(G, D)=\lambda_{ {adv }} L_{ {adv }}(G, D) \\
&+E_{ {v} \sim p_{ {data }}, z} L_{ {aux}}\left(v, \hat{v}\right)
\end{aligned}
\end{equation}
where $v$ is the original audio while $\hat{v} = G(z, {Y}_{1: M})$ is the predicted audio, $p_{data}$ represents the distribution of ground-truth waveform data, $z$ represents our injected Gaussian noise, and $\lambda_{adv}$ is a tunable parameter to balance the performance between tasks.

On the other hand, loss equation of the discriminator defined below aims at strengthening its ability to tell the generated waveforms from the ground-truth:

\begin{equation}
\begin{aligned}
L_{D}(G, D) &=E_{ {v} \sim p_{ {data }}}\left[\left(1-D\left( {v}\right)\right)^{2}\right] \\
&+E_{z}\left[D(G(z, {Y}_{1: M}))^{2}\right]
\end{aligned}
\end{equation}

The block diagram of PWG is shown in Fig.~\ref{fig:PWG}. 
The generator and discriminator are optimized according to a certain strategy during the training stage, and the trained generator is further used in the inference stage to produce the final results of the SSRNet.

\begin{table}[]
\caption{Hyper-parameter of SSRNet}
\centering
\label{Hyper:SSR}
\begin{tabular}{ccc}
\hline
\multicolumn{2}{c}{Item} & Details \\ \hline
\multicolumn{2}{c}{attention transformation dimensions} & 384 \\ \hline
\multicolumn{2}{c}{heads for multi head attention} & 4 \\ \hline
\multirow{2}{*}{source encoder} & FFT layers & 6 \\
 & hidden units & 1536 \\ \hline
\multirow{2}{*}{target decoder} & FFT layers & 6 \\
 & hidden units & 1536 \\ \hline
\multirow{3}{*}{postnet} & layers & 5 \\
 & filter channels & 256 \\
 & filter size & 5 \\ \hline
\multirow{3}{*}{duration predictor} & layers & 2 \\
 & filter channels & 384 \\
 & kernel size & 3 \\ \hline
\end{tabular}
\end{table}

\begin{table}[!ht]
\caption{Hyper-parameter of PWG}
\centering
\label{Hyperparameter}
  \begin{tabular}{ c  c  c }
  \hline
    Item & \multicolumn{2}{c}{Objects and/or Details} \\ \hline
    $\lambda_{adv}$ & \multicolumn{2}{c}{4}\\
    filter size & \multicolumn{2}{c}{3} \\
    batch size & \multicolumn{2}{c}{6} \\
    training audio length & \multicolumn{2}{c}{16384 (1.024s)} \\
    WaveNet generator & \multicolumn{2}{c}{30-layer dilated residual convolution } \\
    discriminator & \multicolumn{2}{c}{10-layer dilated residual convolution } \\\hline
    \multirow{2}{*}{learning rate} &   generator & 1e-4  \\
    & discriminator & 5e-5 \\ \hline
    \multirow{2}{*}{training steps} & generator-only & 1e6 \\
    & jointly & 4e6 \\ \hline
    \multirow{2}{*}{channel size} & skip channels & 64 \\
    & residual channels & 64 \\ \hline
    optimizer & RAdam optimizer & $\epsilon=1e-6$ \\
    activation function & Leaky ReLU & $\alpha=0.2$ \\
    \hline
\end{tabular}
\end{table}

\section{Experiments and Results}\label{sec:results}
In this section, we first introduce the experimental setting. Then, we evaluate experimental results with the objective and subjective metrics and compare them with the baseline model. Furthermore, we conduct an ablation study on the model modules. Finally, we provide more insights into the toneme study from the aforementioned results.

\subsection{Experimental Setting}

\begin{table*}[!ht]
\caption{Subjective Comparison between Reconstructed Voices from the Baseline and the SSRNet}\label{tab:human}
\centerline{
\begin{tabular}{ccccccc}

\hline
 & Spk-1 & Spk-2 & Spk-3 & Spk-4 & Spk-5 & Spk-6 \\\hline
Baseline CER(\%) & 55.06$\pm$41.62 & 17.72$\pm$13.96 & 23.00$\pm$2.71 & 53.37$\pm$6.67 & 26.37$\pm$21.18 & 63.05$\pm$27.00 \\
SSRNet CER(\%) & 1.70$\pm$3.4 & 1.19$\pm$1.46 & 2.31$\pm$2.37 & 8.92$\pm$5.77 & 20.67$\pm$5.69 & 3.69$\pm$3.05 \\\hline
Baseline Naturalness & 44$\pm$16 & 50$\pm$14 & 51$\pm$17 & 39$\pm$13 & 51$\pm$20 & 41$\pm$18 \\
SSRNet Naturalness & 95$\pm$7 & 71$\pm$17 & 89$\pm$5 & 64$\pm$10 & 58$\pm$11 & 77$\pm$16\\\hline
\end{tabular}}
\end{table*}

In the training stage of the SSRNet, the batch size is set to 8 utterances. Besides, dropout rate for encoder and decoder is set to 0.1 and for postnet is set to 0.5. The detailed settings of SSRNet are shown in Table~\ref{Hyper:SSR}. The Adam optimization algorithm is used to optimize trainable parameters. The Noam learning rate (LR) scheduler is used in the training procedure as follows~\cite{DBLP:conf/nips/VaswaniSPUJGKP17}:
\begin{equation}
lr\ = d_{model}^{-0.5}*\min(step^{-0.5},step*step_{w}^{-1.5})
\end{equation}
where ${step_{w}}$ is set to 4000, $d_{model}$ is set to 384, and $step$ denotes the number of the training step. These parameter values are chosen based on~\cite{DBLP:conf/nips/VaswaniSPUJGKP17}. Furthermore, $\lambda_{{align}}$ in Eq.~(\ref{eq:align}) is set to 10, $\lambda_{tm}$ and $\lambda_{recons}$ in the Eq.~(\ref{eq:loss}) are both set to 0.5. 
The GT duration of the training set is calculated as Eq.~(\ref{eq:align_old}) and Eq.~(\ref{eq:dura}) before training. And the model uses these initial GT duration to calculate the loss in the first four epochs.
In the training stage, the GT duration is updated every five epochs by Eq.~(\ref{eq:align}) and Eq.~(\ref{eq:dura}). The implementation of the SSRNet model is based on the ESPNET toolkit\footnote{https://github.com/espnet/espnet}~\cite{DBLP:conf/interspeech/WatanabeHKHNUSH18}. 


For vocoder, PWG is pre-trained within $Audio_{v}$ of multi speakers in the training set. In the first 100K steps of the training stage, the discriminator parameters are fixed, and only the generator is trained on the first stage. After that, the two modules are jointly trained until 400K steps to further build the synthesis quality. Our experiment is based on the implementation of PWG\footnote{https://github.com/kan-bayashi/ParallelWaveGAN}. The detailed settings of PWG are shown in Table~\ref{Hyperparameter}.

\begin{figure}[!t]
\centerline{
\includegraphics[width=3.3in]{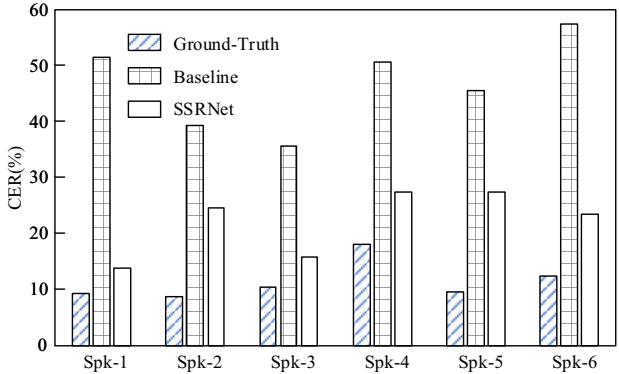}}
\caption{Objective accuracy comparison between the ground-truth voices, reconstructed voices from the baseline and the SSRNet.}
\label{fig:accuracy}
\end{figure}

The previous work proposed by Gaddy and Klevin~\cite{DBLP:conf/emnlp/GaddyK20} is considered as the baseline model. The training parameters of the baseline model are consistent with those reported in~\cite{DBLP:conf/emnlp/GaddyK20}. The training, validation, and testing data used are the same as those used in the SSRNet model. Moreover, we employ the pre-trained PWG instead of WaveNet as the vocoder in baseline to deal with the limitation of inference speed~\cite{DBLP:conf/icassp/YamamotoSK20}. We train the SSRNet and baseline separately for each participant.


\subsection{Model Performance on the $sEMG\_Mandarin$ Dataset}
\subsubsection{Objective Evaluation}

The objective evaluation is about the quality and accuracy of reconstructed voices. For the objective accuracy evaluation, this paper employs an automatic speech recognition (ASR), called Mandarin ASR (MASR)\footnote{https://github.com/nobody132/masr}, as metrics. MASR uses the character error rate (CER) with the Levenshtein distance to measure the accuracy between the predicted text and the original text~\cite{DBLP:journals/csur/Navarro01}. Note that CER ranges from 0 to +$\infty$.  
CER can get infinite because the ASR can insert an arbitrary amount of words~\cite{DBLP:conf/icnlsp/ErrattahiHO15}.
In the experiments, CER based on ASR for each epoch is calculated on the validation set by the model, and parameters of the best CER epoch is selected as the best-performed final model. 

It is observed in Fig.~\ref{fig:accuracy} that the proposed method SSRNet outperforms the baseline significantly for all six speakers. The SSRNet obtains an average CER of 21.99\% in ASR with a standard deviation of 4.99\% across six speakers. Besides, The SSRNet outperforms the baseline by 24.63\%. Meanwhile, the ground-truth voices from the testing set achieve a CER of 11.30\%. It verifies that SSRNet generates more intelligible voices. Because SSRNet calculates the duration of the silent speech, regulates the silent sEMG following audio length, and uses a multi-task learning strategy to improve results. Besides, the results across speakers differ, among which the worst accuracy is achieved on Spk-4 with a CER of 27.20\% and Spk-5 with a CER of 27.34\% the best accuracy is achieved on Spk-1 with a CER of 13.62\%. 
By studying the two speaker cases with the worst accuracy, we find that the ground-truth voices of Spk-4, which performs poorly on ASR, can cause low accuracy. At the same time,  higher impedance resulting in a lower signal-to-noise ratio during the experiment leads to the wrong result on Spk-5.




For the objective quality evaluation, we use Mel-Cepstral Distortion (MCD)\footnote{https://github.com/mpariente/pystoi}~\cite{kubichek1993mel} and Short Term Objective Intelligibility (STOI)\footnote{https://github.com/ttslr/python-MCD}~\cite{DBLP:conf/icassp/TaalHHJ10}. The lower MCD indicates a higher similarity between the synthesized and the natural mel-cepstral sequences. Meanwhile, the higher STOI reflects higher intelligibility and better clarity of the speech.

\begin{figure}[!t]
\centerline{
\includegraphics[width=3.3in]{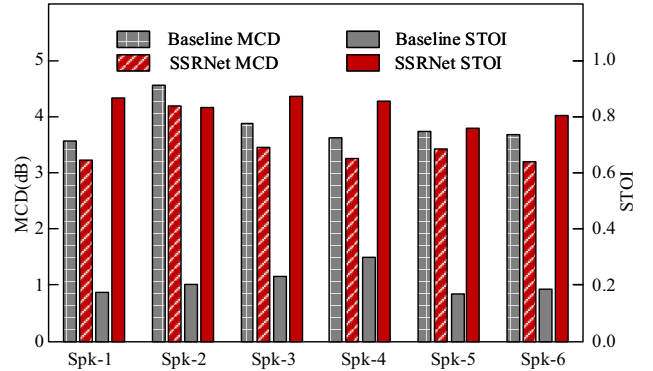}}
\caption{Objective quality comparison between reconstructed voices from the baseline and the SSRNet.}
\label{fig:quality}
\end{figure}

Fig.~\ref{fig:quality} summarizes the MCD and STOI evaluation. 
It is observed that the SSRNet model consistently performs better than the baseline model for both quality and intelligibility.
The reason is that the length of reconstructed voice in the baseline is consistent with silent speech and impaired. As a comparison, SSRNet firstly provides length-regulated voices, which are more similar to the ground-truth voices.


\begin{table*}[!t]
\caption{Ablation Results on the Model Module Study}
\label{Tab:ablation}
\centerline{
\begin{tabular}{cccccc}
\hline
\makecell[c]{$sEMG_{v}$\\Reconstruction Module} & \makecell[c]{Toneme Classification\\Module}& \makecell[c]{Toneme Classification\\Module Position}& \makecell[c]{Tones in Toneme\\Classification Module} &\makecell[c]{The Cost Function\\for DTW}& \makecell[c]{$\Delta$CER(\%)}\\\hline

$\checkmark$ ($\lambda_{\text {recons}}$=0.5)            & $\checkmark$ ($\lambda_{\text {tm}}$=0.5)                         & Before Decoder    &$\checkmark$  &\makecell[c]{$\lambda_{\text {align}}$=10} & +0  \\
\hline

\boldmath$\boldtimes$ ($\lambda_{\text {recons}}$=0)  \unboldmath          & $\checkmark$ ($\lambda_{\text {tm}}$=0.5)                       & Before Decoder   &$\checkmark$ &\makecell[c]{$\lambda_{\text {align}}$=10}   & +9.38      \\\hline
$\checkmark$ ($\lambda_{\text {recons}}$=0.5)           &\textbf{$\boldtimes$ ($\lambda_{\text {tm}}$=0)}                         & -           &$\checkmark$         &\makecell[c]{$\lambda_{\text {align}}$=10}         & +132.75\\\hline
$\checkmark$ ($\lambda_{\text {recons}}$=0.5)            & $\checkmark$ ($\lambda_{\text {tm}}$=0.5)                        & \textbf{After Decoder}        &$\checkmark$    &\makecell[c]{$\lambda_{\text {align}}$=10}    & +1.89
 
     \\\hline
$\checkmark$ ($\lambda_{\text {recons}}$=0.5)            & $\checkmark$ ($\lambda_{\text {tm}}$=0.5)                         & Before Decoder    &$\boldtimes$  &\makecell[c]{$\lambda_{\text {align}}$=10} & +6.51  \\
\hline
$\checkmark$ ($\lambda_{\text {recons}}$=0.5)            & $\checkmark$ ($\lambda_{\text {tm}}$=0.5)                         & Before Decoder     &$\checkmark$ &\makecell[c]{\textbf{$\lambda_{\text {align}}$=0}} & +81.18
     \\\hline

\end{tabular}}
\end{table*}

\subsubsection{Subjective Evaluation}
We use subjective evaluation based on the transcriptions from 10 native Mandarin Chinese human listeners. The average age of the ten listeners is 24. These listeners have no prior knowledge of the context of the voices. They are required to listen to the voices with earphones in a quiet environment. Each listener is required to listen to 60 sample voices from 6 speakers, randomly selected from the SSRNet and baseline testing set. They are asked to transcribe the audios into text in Mandarin Chinese and give a score of the naturalness of each speaker ranging from 0 to 100 (0 for the worst naturalness while 100 for the best). 

\textbf{The reconstructed samples can be found on our website\footnote{https://irislhy.github.io/}}. 

The results of human evaluation of six-speakers samples are shown in Table~\ref{tab:human} and $\pm$ indicates the standard deviation of the metrics across evaluation of listeners. The results of the subjective evaluation are consistent with the objective evaluation. The SSRNet gets an average CER of 6.41\%, while the baseline gets an average CER of 39.76\% in subjective evaluation. Besides, the naturalness scores from listeners are consistent with the objective evaluation results.  
Our exploratory analysis shows that the proposed SSRNet outperforms the baseline in human intelligibility and naturalness.

 \begin{figure}[!t]
\centerline{
\includegraphics[width=3.3in]{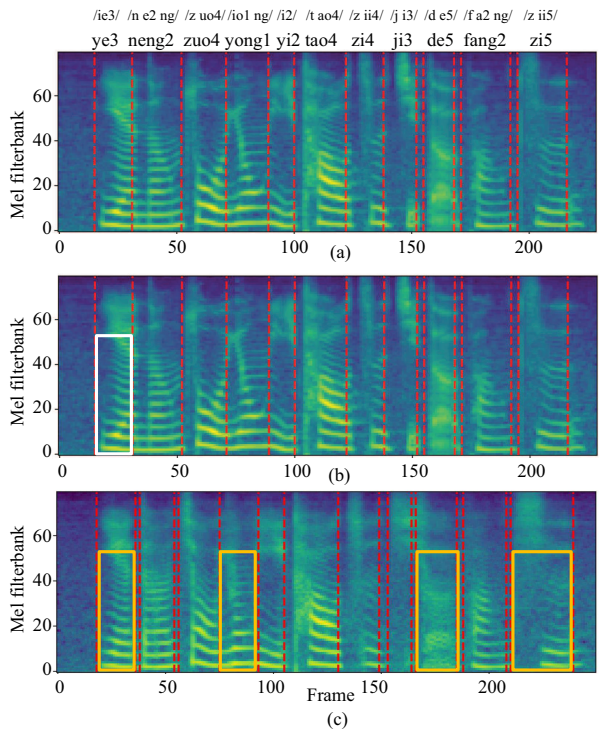}}
\caption{Mel-spectrograms visualizations of (a) Ground-truth voice, (b) Voice reconstructed by the SSRNet, (c) Voice reconstructed by the baseline.
}
\label{fig:spec_compare}
\end{figure}

\subsubsection{Mel-spectrograms Comparison}

Fig.~\ref{fig:spec_compare} depicts mel-spectrograms of one example from the testing set of Spk-3. We have three observations, 1) The mel-spectrograms synthesized by the SSRNet is more close to the ground-truth one and has a similar length. This is because that the length regulator resamples the length of the output frames. Based on this, SSRNet deals with the lack of time-aligned data of vocal and silent speech, and generates more natural sounds. 2) The white color box indicates a slight blurring of the pronunciation /ye3/ in SSRNet compared to the ground-truth pronunciation, but listeners and ASR can understand it. 3) The yellow color boxes indicate four examples of errors for the baseline that some listeners and ASR have difficulty understanding, and the voice synthesized from the baseline is not clear overall. 

In conclusion, the experiments demonstrate that SSRNet provides a solution to narrow the gap between the reconstructed and natural voices.


\subsection{Ablation Study}\label{sec:ablation}

Next, we conduct ablation studies to gauge the effectiveness of every extension in the SSRNet, including joint optimization, model prediction alignments, and tone evaluation. Because of the consistency between the objective and subjective evaluation, only the objective accuracy evaluation is performed for ablation studies. Table~\ref{Tab:ablation} summarizes the ablation study results of different model modules. The first row shows the settings of SSRNet, while the final column shows the change in average CER across six speakers compared to the SSRNet.


\subsubsection{Joint Optimization} 
The second to third rows show the changes in consequence of removing the joint optimization. It is observed that removing the joint optimization could lead to performance degradation in terms of accuracy. This indicates that the toneme classification and the $sEMG_{v}$ reconstruction are practical for the SSRNet. Note that the toneme classification module contributes significantly more in the SSRNet than $sEMG_{v}$ reconstruction. We find that removing the toneme classification results in an absolute difference between the context of synthesized voices and the ground-truth context. It means in the Seq2Seq model, the hidden representations after the length regulator have difficulty in obtaining the context information of the $sEMG_{s}$. As a result, joint optimization is conducive to studying feature transformation.


\begin{figure*}[!t]
\centerline{
\subfloat[Consonants confusability]{
     \label{fig:Consonants}
    \includegraphics[height=2.3in]{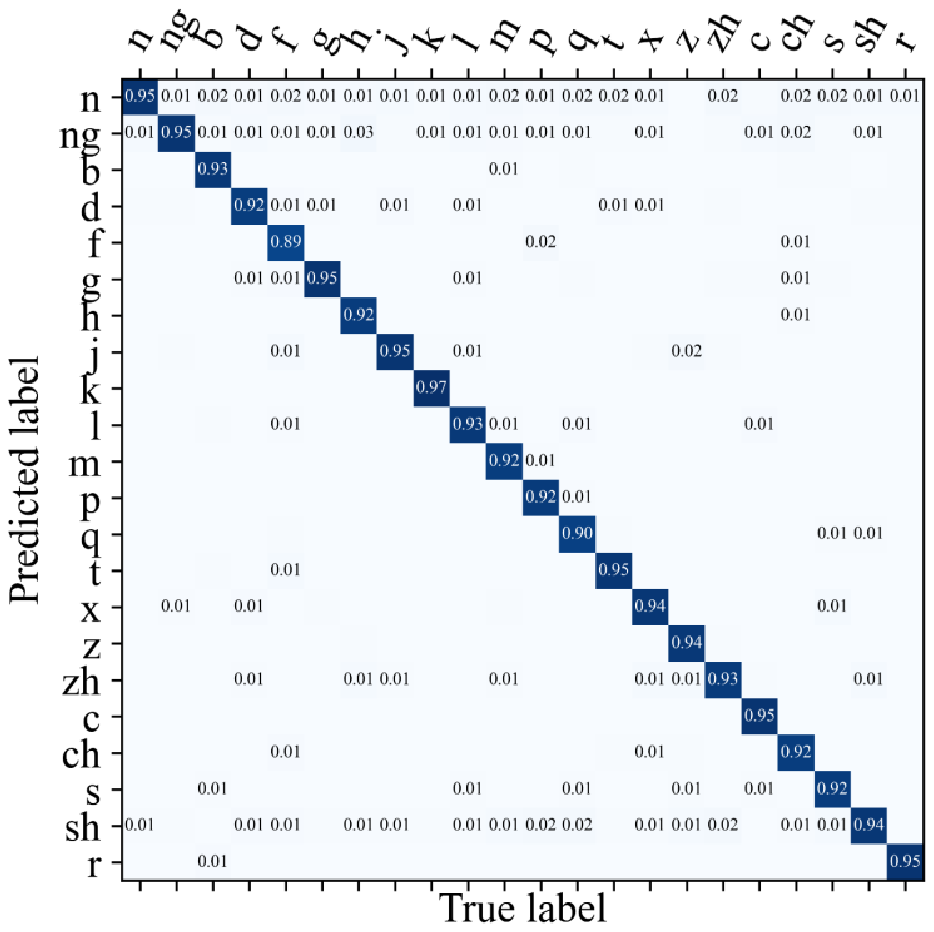}
    }
\subfloat[Vowels confusability]{
     \label{fig:vowels}
    \includegraphics[height=2.3in]{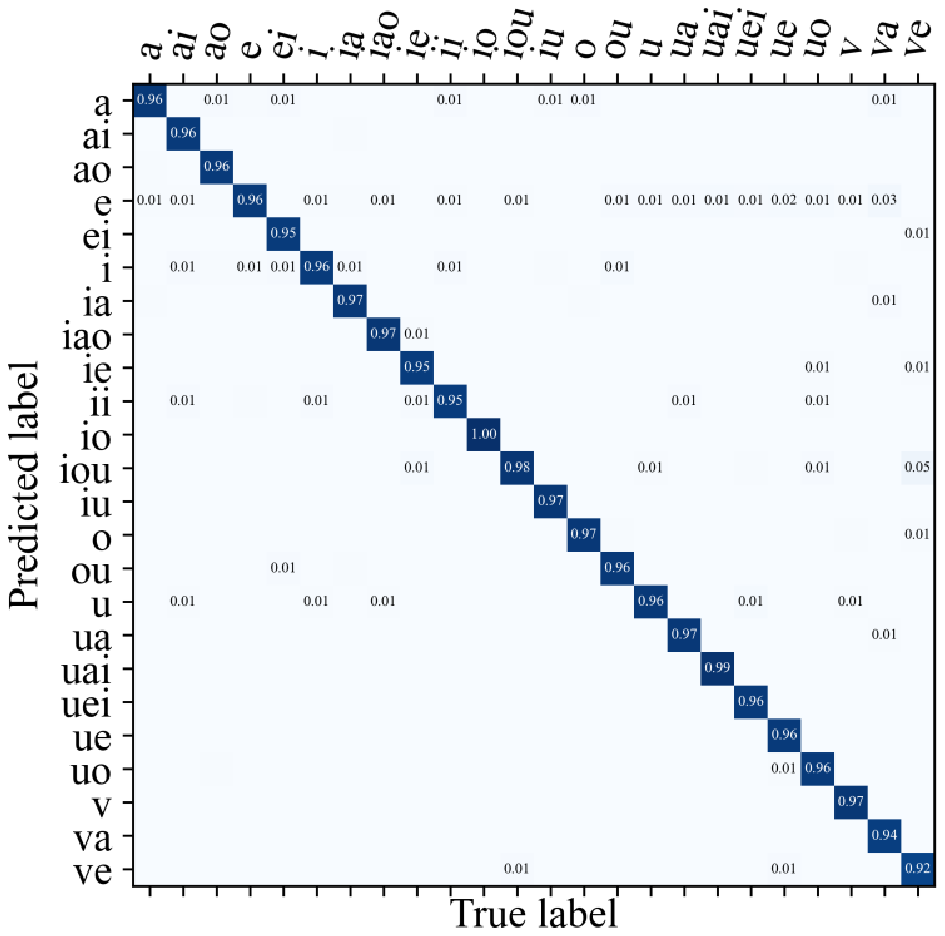}
    }
\subfloat[Tones confusability]{
     \label{fig:tones}
    \includegraphics[height=2.3in]{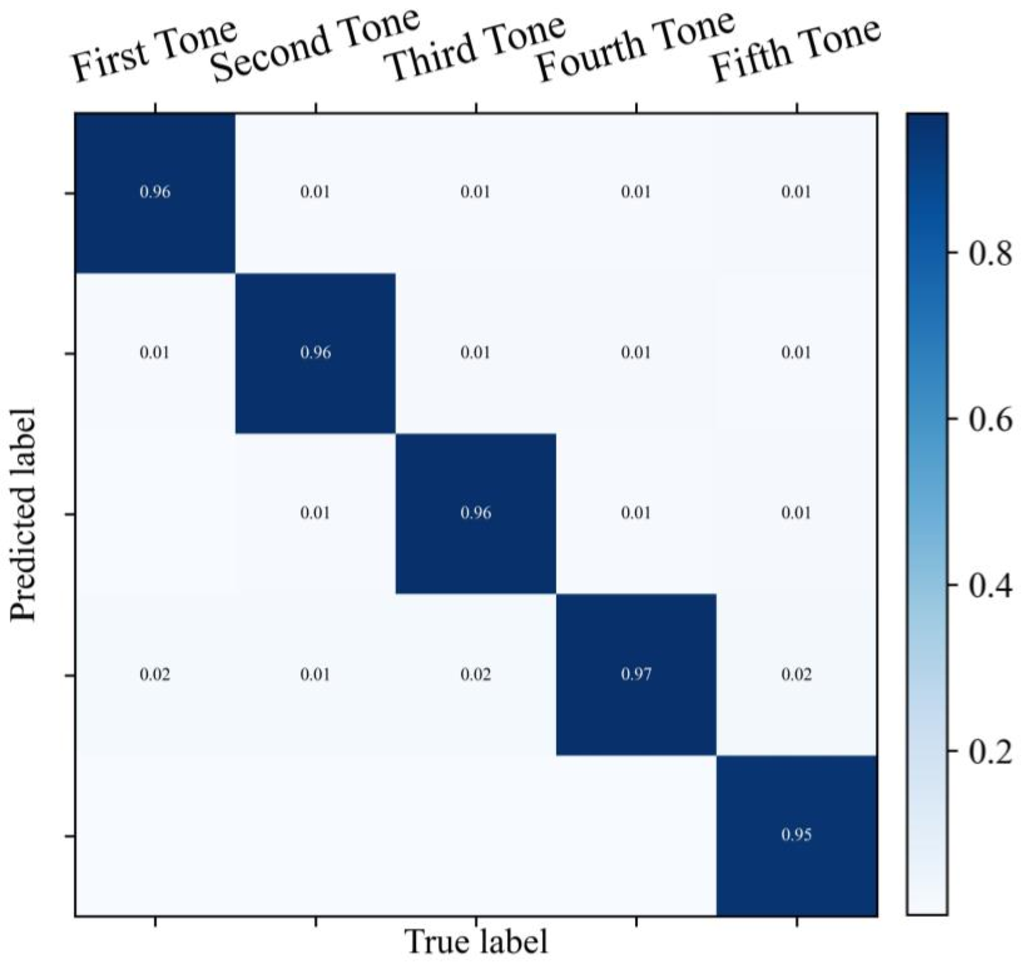} 
}}
\caption{Tonemes confusion matrices on the testing set, number in row j, column i is the ratio of the number of samples predicted as label j with true label i to number of samples with true label i. Values smaller than 0.005 are ignored.}
\label{Fig:confusion}
\end{figure*}
\subsubsection{The Position of the Toneme Classification Module}
We also investigate the position of the toneme classification module by comparing results in the fourth row with those in the first row while the position of $sEMG_{v}$ reconstruction is fixed. In the fourth row of the table, the position of the toneme classification is located after the decoder. In contrast, the position in the first row is located before the decoder. The position before the decoder outperforms the position after decoder by 1.89\%. This implies that the position of the module in the middle layer or final layer can both represent the source content in the $sEMG2V$ task.

\subsubsection{Tone in Toneme Classification} 
We also conduct the tones evaluation. We use phoneme classification instead of toneme classification. The phoneme classification module predicts a sequence and measures the CE loss between true and predicted phonemes without any tone information. We find that lack of tone information resulted in a 6.51\% increase in CER in the fifth row , which demonstrates that the $sEMG2V$ task in Mandarin Chinese needs tone information in concert with phoneme rather than separate phoneme information.

\subsubsection{The Cost Function for DTW} 

We conduct the alignment study as described in the sixth row. It shows that the CER of the alignments strategy in SSRNet shows a relative reduction of over 81.18\% compared to the traditional approach, which demonstrates that effectiveness of the alignments between $N$-length predicted audio features ${\hat Y}^{+*}_{1: N}$ obtained by the SSRNet model without length regulator and $M$-length ground-truth audio features ${Y}_{1: M}$.

\subsection{Frame-based Toneme Classification Study}
At last, we evaluate the frame-based performance of the toneme classification module on the testing set except for silent frames. We use the GT-duration calculated by Eq.~(\ref{eq:dura}) with the best-performed model of each speaker to match the length of ground-truth phonemes. Because the confusion between vowel consonants is interpretable~\cite{DBLP:journals/corr/abs-2106-01933}, this section focuses on the vowel pairs, consonants pairs, and tone pairs. 
The confusion matrices are calculated to elaborate more toneme prediction details, as shown in Fig.~\ref{Fig:confusion}. 

It can be seen in Fig.~\ref{Fig:confusion}(a) and Fig.~\ref{Fig:confusion}(b) that SSRNet provides excellent classification results for consonants and vowels. We observe the confusion between nasal and other consonants, which is consistent with~\cite{freitas2015detecting} and~\cite{DBLP:journals/corr/abs-2106-01933}. This is due to the limitations of sEMG electrodes in detecting velum~\cite{freitas2015detecting}.

Meanwhile, Fig.~\ref{Fig:confusion}(c) shows the confusion matrix of the tone set, which is calculated from the ground-truth tones and the predicted tones from vowels and is directly extracted from the entire confusion matrix. The tone classification achieves an average accuracy of 96.07\%. This proves that neuromuscular signals can transfer most tone information in silent speech. The fifth tone is sometimes mistaken as the other four tones. This indicates that the fifth tone is sometimes difficult to express in silent speech. 


\section{Conclusion}\label{sec:conclusion}
This paper proposes a Seq2Seq-based model SSRNet to decode neuromuscular signals in a tonal language. SSRNet uses the duration extracted from the alignment to regulate the sEMG-based silent speech. Furthermore, a toneme classification module and a vocal sEMG reconstruction module are used to improve the overall performance. We conduct extensive experiments on the Mandarin Chinese dataset to demonstrate that the proposed model outperforms the baseline model in both objective and subjective evaluation. The model achieves an average subjective CER of 6.41\% for six speakers and 1.19\% for the best speaker, demonstrating the feasibility of the reconstruction task. 

In the future, we would like to enhance the robustness of the model by including more speakers. Another possible direction is making the system real-time because it is necessary for speakers to learn to improve pronunciation by themselves in silent speech based on auditory feedback. 

\appendices

\bibliographystyle{IEEEtran}
\bibliography{IEEEabrv,mybib}

\end{document}